\documentclass[prd,aps,nofootinbib,twocolumn,showpacs,floats,letterpaper,floatfix,groupedaddress,eqsecnum]{revtex4}

\usepackage{dcolumn,epsfig}
\usepackage{amssymb,amsmath}

\def\t{\times}
\def\be{\begin{equation}}
\def\ee{\end{equation}}
\def\beq{\begin{eqnarray}}
\def\eeq{\end{eqnarray}}

\begin{document}

\title{Ergoregion instability of ultra-compact astrophysical objects}

\author{Vitor Cardoso} \email{vcardoso@phy.olemiss.edu}
\affiliation{Department of Physics and Astronomy, The University
of Mississippi, University, MS 38677-1848, USA \footnote{Presently
at CENTRA, Dept. de F\'{\i}sica, Instituto Superior T\'ecnico, Av.
Rovisco Pais 1, 1049-001 Lisboa, Portugal}}

\author{Paolo Pani} \email{paolo.pani@ca.infn.it}
\affiliation{Dipartimento di Fisica, Universit\`a di Cagliari, Cittadella Universitaria 09042 Monserrato, Italy
\\ and Department of Physics and Astronomy, The University of Mississippi, University, MS 38677-1848, USA}

\author{Mariano Cadoni} \email{mariano.cadoni@ca.infn.it}
\affiliation{Dipartimento di Fisica, Universit\`a di Cagliari, and
INFN sezione di Cagliari, Cittadella Universitaria 09042 Monserrato,
Italy}

\author{Marco Cavagli\`a} \email{cavaglia@phy.olemiss.edu}
\affiliation{Department of Physics and Astronomy, The University of Mississippi, University, MS 38677-1848,
USA}

\date{\today}

\begin{abstract}
Most of the properties of black holes can be mimicked by horizonless compact objects such as
gravastars and boson stars. We show that these ultra-compact objects develop a strong
ergoregion instability when rapidly spinning. Instability timescales can be of the order of 0.1
seconds to 1 week for objects with mass $M=1-10^6M_{\odot}$ and angular momentum $J> 0.4 M^2$.
This provides a strong indication that ultra-compact objects with large rotation are black
holes. Explosive events due to ergoregion instability have a well-defined gravitational-wave
signature. These events could be detected by next-generation gravitational-wave detectors such
as Advanced LIGO or LISA.
\end{abstract}


\maketitle
\section{Introduction}
Black holes (BHs) in Einstein-Maxwell theory are characterized by three parameters
\cite{Hawking:1971vc}: Mass $M$, electric charge $Q$ and angular momentum $J \equiv a M
\leqslant M^2$. BHs are thought to be abundant objects in the Universe \cite{Narayan:2005ie}.
Their mass is estimated to vary between $3 M_{\odot}$ and $10^{9.5} M_{\odot}$ or higher. They
are likely to be electrically neutral because of the effect of surrounding plasma
\cite{Blandford:1977ds} and their angular momentum is expected to be close to the extremal
limit because of accretion and merger events \cite{Gammie:2003qi,Merritt:2004gc}. An example of
astrophysical BH is the compact primary of the binary X-ray source GRS $1915+105$, which recent
observations identify as a rapidly-rotating object of spin $a\gtrsim 0.98~M$
\cite{McClintock:2006xd}. Many of the supermassive BHs which are thought to power quasars seem
to be rotating near the Kerr bound \cite{Wang:2006bz}.

Despite the wealth of circumstantial evidence, there is no definite observational proof of the
existence of astrophysical BHs. (A review and a critique of current evidence can be found in
Ref.\ \cite{Narayan:2005ie} and Ref.\ \cite{Abramowicz:2002vt}, respectively. See also Ref.\
\cite{Lasota:2006jh} for a stimulating mini-review.) Astrophysical objects without event
horizon, yet observationally indistinguishable from BHs, cannot be excluded a priori.

Dark energy stars or ``gravastars'' are compact objects with de Sitter interior and
Schwarzschild exterior \cite{Chapline:2000en,Mazur:2001fv}. These two regions are glued
together around the would-be horizon by an ultra-stiff thin shell. In this model, a
gravitationally collapsing star undergoes a phase transition that prevents further collapse.
The thickness of the shell sets an upper limit to the mass of the gravastar \cite{Mazur:2001fv,
Visser:2003ge, Chirenti:2007mk}. (A thorough analysis of the maximum compactness of gravastars
can be found in Ref.\ \cite{Horvat:2007qa}.) Generalizations of the original model use a
Born-Infeld phantom field \cite{Bilic:2005sn}, dark energy equation of state \cite{Lobo:2005uf}
or non-linear electrodynamics \cite{Lobo:2006xt}. Models without shells or discontinuities have
been investigated in Ref.\ \cite{Cattoen:2005he}.

Boson stars are macroscopic quantum states which are prevented from undergoing complete
gravitational collapse by Heisenberg uncertainty principle \cite{bosonstars,Yoshida:1997nd}.
Their models differ in the scalar self-interaction potential \cite{Schunck:2003kk} and can be
divided in three classes \cite{Berti:2006qt}:

\medskip\noindent {\it Miniboson stars.} If the scalar field is non-interacting, the maximum
boson star mass is $M_{\rm max}\sim 0.633 m^2_{\rm Planck}/m$ \cite{bosonstars}. This value is
much smaller than the Chandrasekhar mass for fermion stars, $M_{\rm Ch}\sim m^3_{\rm
Planck}/m^2$. Stability of supermassive objects requires an ultralight boson of mass
$m=8.45\times 10^{-26} {\rm GeV}~(10^6 M_\odot/M_{\rm max})$.

\medskip\noindent {\it Massive boson stars.} The requirement of ultralight bosons can be lifted
if the scalar field possesses a quartic self-interaction potential of the form $\lambda
|\phi|^4/4$ \cite{Colpi:1986ye}. As long as the coupling constant $\lambda$ is much larger than
$(m/m_{\rm Planck})^2$, the maximum boson star mass can be of the order of the Chandrasekhar
mass or larger, $M_{\rm max}\sim 0.062 \lambda^{1/2} m^3_{\rm Planck}/m^2$. Thus supermassive
objects may exist. Boson mass and coupling constant are related by $m=3.2\times 10^{-4}~{\rm
GeV}~\lambda^{1/4} (10^6 M_\odot/M_{\rm max})^{1/2}$.

\medskip\noindent {\it Nontopological soliton stars.} If the self-interaction takes the form
$U=m^2|\phi|^2(1-|\phi|^2/\phi_0^2)^2$, compact nondispersive solutions with a finite mass may
exist even in the absence of gravity \cite{Lee:1991ax}. The critical mass of these objects is
$M_{\rm max}\sim 0.0198m^4_{\rm Planck}/(m \phi_0^2)$. If $\phi_0\sim m$, a star of mass $M\sim
10^6 M_\odot$ corresponds to a heavy boson of mass $m\sim 500$~GeV.

\medskip
Boson stars are indistinguishable from BHs in the Newtonian regime. Since they are very
compact, deviations in the properties of orbiting objects occur close to the Schwarzschild
radius and are not easily detectable electromagnetically \cite{Torres:2000dw,Guzman:2005bs}. If
the scalar field interacts only gravitationally with matter, compact objects may safely
inspiral ``inside'' the boson star, the only difference with a BH being the absence of an event
horizon \cite{Kesden:2004qx}. Lack of strong constraints on boson masses makes these models
difficult to rule out.

Gravastars and boson stars provide viable alternatives to astrophysical BHs. To ascertain the
true nature of ultra-compact objects it is thus important to devise observational tests to
distinguish these ultra-compact objects from ordinary BHs. The traditional way to distinguish a
BH from a neutron star is to measure its mass. If the latter is larger than the Chandrasekhar
limit, the object is believed to be a BH. However, this method cannot be used for the
ultra-compact objects discussed above, because of their broad mass spectrum. A possibility is
to look for observables related to the accretion mechanism. For example, the luminosity of
quiescent BHs is lower than the maximum luminosity which is allowed by the gas present in their
environment \cite{fabian:1988}. If the BH accretion rate is much smaller than the Eddington
rate, the radiative efficiency is also very small \cite{Narayan:1995ic}. Another possibility is
to exploit the absence of a boundary layer at the surface. Compact stars with accretion disks
have typically a narrow viscous boundary layer near their surface, which allows the release of
a considerable amount of heat energy. On the other hand, if the central object is a BH, no
boundary layer is formed. Arguments of this kind have already excluded many gravastar
candidates \cite{Broderick:2007ek}. Absence of type I X-ray bursts is another powerful
indicator of the presence of a BH. Several studies on type I bursts show that they are produced
when gas accretes on the surface of a NS \cite{lewin:1993}, which then undergoes a semi-regular
series of thermonuclear explosions. Since BHs do not have surfaces, the surrounding gas cannot
accumulate and thermonuclear instabilities do not develop.

Another very promising observational method to probe the structure of ultra-compact objects is
gravitational wave astronomy \cite{Schutz:1999xj}. Gravitational wave detectors such as LIGO
\cite{ligo}, VIRGO \cite{virgo}, TAMA \cite{tama}, or LISA \cite{lisa} could provide an
efficient way to study these objects without intervening effects due to the interstellar
medium. For example, the inspiral process of two compact objects allows a precise determination
of their mass \cite{Berti:2004bd} and multipole moments \cite{Kesden:2004qx, Ryan:1995wh,
Barack:2006pq, AmaroSeoane:2007aw}. The gravitational waveform in the presence of a surface is
also expected to be different than the waveform in the presence of an event horizon
\cite{Vallisneri:1999nq}. A first study on the distinctive features of the inspiral signal of
boson stars can be found in Ref.\ \cite{Palenzuela:2007dm}. Detection of gravitational resonant
modes due to the gravitational potential well could also provide a test for the presence of a
horizon \cite{Berti:2006qt,Berti:2005ys}. Preliminary studies for gravastars indicate that this
method may be very efficient if the source is not too far away and gravitational wave
production is significant \cite{Berti:2006qt,Yoshida:1994xi,Chirenti:2007mk}.

In this paper, we propose a new method for discriminating BHs from ultra-compact horizonless
objects and apply it to gravastars and bosons stars. Our method uses the fact that compact
rotating objects without event horizon are unstable when an ergoregion is present. The origin
of this {\it ergoregion instability} can be traced back to superradiant scattering. In a
scattering process, superradiance occurs when scattered waves have amplitudes larger than
incident waves. This leads to extraction of energy from the scattering body
\cite{zel1,staro1,Bekenstein:1998nt}. Instability may arise whenever this process is allowed to
repeat itself ad infinitum. This happens, for example, when a BH is surrounded by a ``mirror''
that scatters the superradiant wave back to the horizon, amplifying it at each scattering. The
total extracted energy grows exponentially with time until the radiation pressure destroys the
mirror in a process called {\it BH bomb} (see Refs.\ \cite{bhbombPress, Cardoso:2004nk}). If
the mirror is inside the ergoregion, superradiance may lead to an inverted BH bomb. Some
superradiant waves escape to infinity carrying positive energy, causing the energy inside the
ergoregion to decrease and eventually generating an instability. This may occur for any
rotating star with an ergoregion: The mirror can be either its surface or, for a star made of
matter non-interacting with the wave, its center. BHs are stable, which could be due to the
absorption by the event horizon being larger than superradiant amplification.

The ergoregion instability appears in any system with ergoregions and no horizons
\cite{friedman}. (See also Ref.~\cite{Vilenkin:1978uc} for a exhaustive discussion.) Explicit
computations for ordinary rotating stars can be found in Ref.\ \cite{cominsschutz,eriguchi},
where typical instability timescales are shown to be larger than the Hubble time. In this case,
the ergoregion instability is too weak to produce any effect on the evolution of the star. This
conclusion changes drastically for ultra-compact stars. For compactness $M\gtrsim 0.5 R$ and
angular momentum $J\gtrsim 0.4M^2$, we find that instability timescales range approximately
from $0.1$ seconds to 1 week for objects with mass in the range $M\sim 1M_{\odot}$ to
$10^6M_{\odot}$, further decreasing for larger rotation rates.

Due to the difficulty of handling gravitational perturbations for rotating objects, the
calculations below are mostly restricted to scalar perturbations. However, we are able to show
that the equation for axial gravitational perturbations of gravastars is identical to the
equation for scalar perturbations in the large $l=m$ limit. There are also generic arguments
suggesting that the timescale of gravitational perturbations is smaller than the timescale of
scalar perturbations for low $m$. Thus our investigation seems to rule out some of these
ultra-compact, rapidly spinning objects as BH candidates.

This paper is organized as follows. In Section \ref{sec:structure} we review the main
characteristics of the ultra-compact objects discussed above. Our discussion is non-exhaustive
and strictly limited to concepts and tools which will be needed in the rest of the paper.
Section \ref{subsec:gravastar} introduces the two gravastar models which will be discussed in
the subsequent analysis. Since there are no known solutions describing rotating gravastars, the
formalism of Refs.\ \cite{Hartle:1967he,schutzexistenceergoregion} will be used to discuss
rotating gravastars. Section \ref{subsec:bosonstars} introduces boson stars
\cite{Colpi:1986ye}. Numerical results for rotating boson stars are taken from Ref.\
\cite{Kleihaus:2005me}. Section \ref{sec:ergoinst} presents a detailed investigation of the
instability of boson stars and gravastars using the WKB approximation. The WKB analysis is then
compared with full numerical results obtained by direct integration of the Klein-Gordon
equation. Detectability of the ergoregion instability by gravitational-wave detectors is
addressed in Sect.\ \ref{sec:grwaves}. Section \ref{sec:discussion} contains a brief discussion
of the results and concludes the paper.

Geometrized units ($G=c=1$) are used throughout the paper, except when numerical results for
rotating boson stars from Ref.\ \cite{Kleihaus:2005me} are discussed (Section
\ref{subsec:bosonstars}). In this case, the Newton constant is defined as $G=0.05/(4\pi)$.

\section{\label{sec:structure} Structure of ultra-compact astrophysical objects}
This section discusses the main properties of gravastars and boson stars.
The derivation of nonrotating solutions is partly based on Refs.\ \cite{Mazur:2001fv,
Chirenti:2007mk,Colpi:1986ye,Kleihaus:2005me}.
\subsection{\label{subsec:gravastar} Gravastars}
Although exact solutions for spinning gravastars are not known, they can be studied in the
limit of slow rotation by perturbing the nonrotating solutions \cite{Hartle:1967he}. This
procedure was used in Ref.~\cite{schutzexistenceergoregion} to study the existence of
ergoregions for ordinary rotating stars with uniform density. Their analysis is repeated below
for gravastars. In the following, we discuss the original thin-shell model by Mazur and Mottola
\cite{Mazur:2001fv} and the anisotropic fluid model by Chirenti and Rezzolla
\cite{Chirenti:2007mk,Cattoen:2005he}.
\subsubsection{Nonrotating thin-shell model
\label{sec:grava1}}
\begin{center}
\begin{figure}[ht]
\begin{tabular}{c}
\epsfig{file=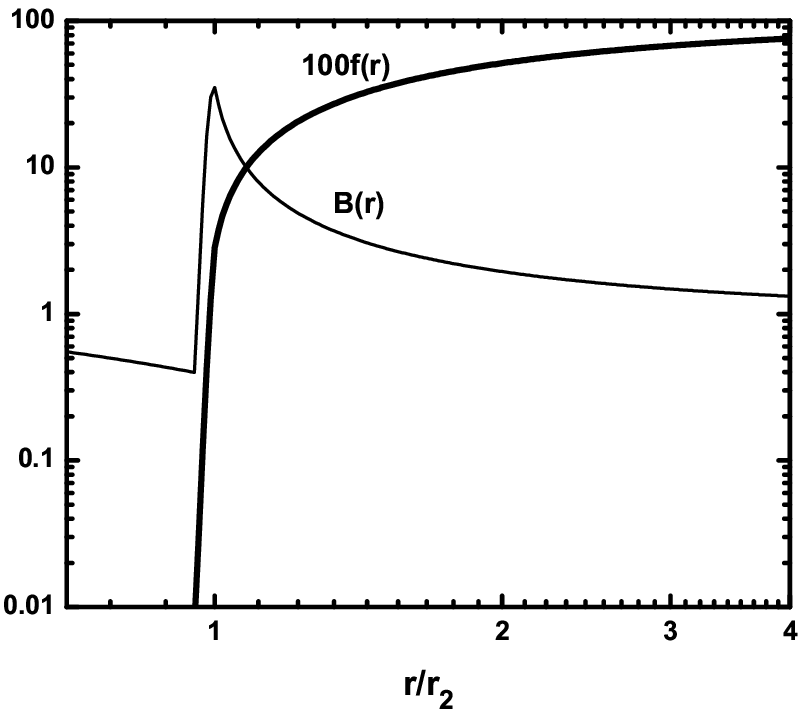,width=200pt,angle=0}\\
\epsfig{file=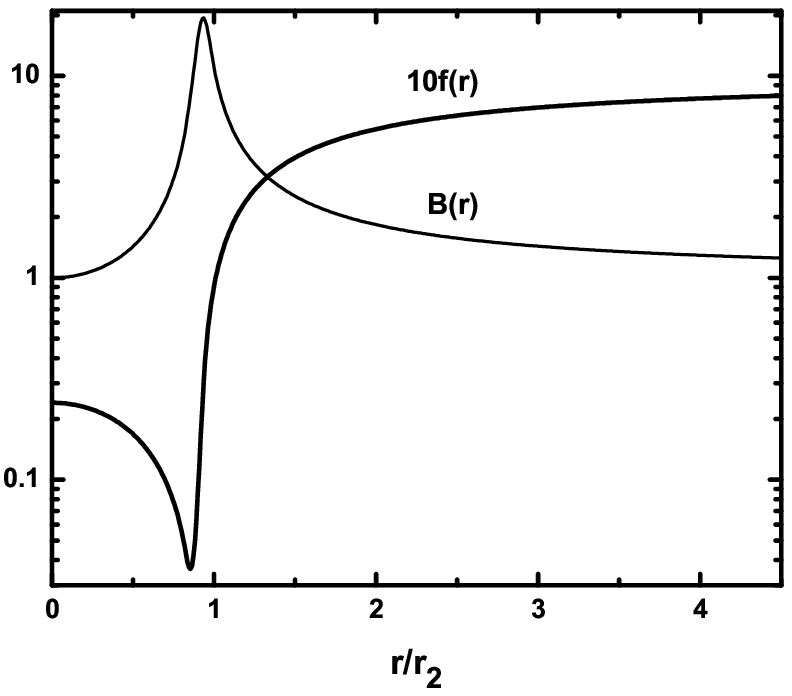,width=200pt,angle=0}
\end{tabular}
\caption{Top panel: Metric coefficients for the thin-shell model gravastar with $r_2=1.05$,
$r_1=1$, $w_1=350$ and $w_2=1$, corresponding to $M\sim 0.485\,r_2$. Bottom panel: Anisotropic
pressure model for $r_2=2.2$, $r_1=1.8$ and $M=1$, corresponding to $M\sim 0.45r_2$.}
\label{fig:metricgrava}
\end{figure}
\end{center}

In this model, the spacetime
\be
ds^2=-f(r)dt^2+B(r)dr^2+r^2d\Omega_2^2
\label{metricspherical}
\ee
consists of three regions:
\begin{eqnarray}
\begin{array}{rlcl}
\textrm{I.} & \textrm{Interior:}\; & 0 \le r \le r_1\,,\; & \rho =-p\,,\\ \\
\textrm{II.} & \textrm{Shell:}\; & r_1 \le r \le r_2\,,\; & \rho =p\,,\\ \\
\textrm{III.} & \textrm{Exterior:}\; & r_2 \le r\,,\; & \rho = p = 0\,,
  \end{array}
\end{eqnarray}
where $\rho$ is the energy density and $p$ is the isotropic pressure of the gravastar. In
region I, $\rho = 3H_0^2/8\pi$ is constant and the metric is de Sitter:
\be
f = \frac{C}{B} = C(1-H_0^2r^2)\,,\qquad 0 \le r \le r_1\,,
\ee
where $C$ is an integration constant to be determined from matching conditions. In region III
the spacetime is described by the Schwarzschild metric,
\begin{equation}
f=\frac{1}{B}=1-\frac{2M}{r}\,,\qquad r_2 \le r\,.
\end{equation}
In region II, the metric is determined by the system of equations,
\beq
d\ln\,r&=&\frac{dh}{1-w-h} \,,\quad h\equiv \frac{1}{B}\label{eqw}\\
d\ln\,h&=&-\left (\frac{1-w-h}{1+w-3h} \right )d\ln\,w\,, \eeq
where $w=8\pi r^2p$ and $wf/r^2=$~constant. A simple analytical solution can be obtained for a
thin shell \cite{Mazur:2001fv}. In the limit $r_1\to r_2$, one obtains
\begin{equation}
\frac{1}{B} \simeq \epsilon\frac{(1+w)^2}{w} \ll 1\,,
\end{equation}
where $\epsilon$ is an integration constant. The continuity of the metric coefficients $f$ and
$B$ at $r_1$ and $r_2$ implies that $\epsilon$, $C$, $M$ and $H_0$ are related to $r_1$, $r_2$,
$w_1\equiv w(r_1)$ and $w_2\equiv w(r_2)$ by \cite{Chirenti:2007mk}
\begin{eqnarray}
\label{int_const} \epsilon &=& -\ln \frac{r_2}{r_1}\left(\ln
\frac{w_2}{w_1}-\frac{1}{w_2}+
\frac{1}{w_1}\right)^{-1}\,,\\
C &=& \left(\frac{1+w_2}{1+w_1}\right)^2\,,\\
 M &=& \frac{r_2}{2}\left[1-\frac{\epsilon (1+w_2)^2}{w_2}\right]\,,\\
 H_0^2 &=& \frac{1}{r_1^2}
 \left[1-\frac{\epsilon (1+w_1)^2}{w_1}\right]\,.
\end{eqnarray}
The above relations and Eq.~(\ref{eqw}) completely determine the structure of the gravastar. A
typical solution is shown in the upper panel of Fig.~\ref{fig:metricgrava} for $r_2=1.05$,
$r_1=1$, $w_1=350$ and $w_2=1$.
%
%
\subsubsection{Nonrotating gravastars with anisotropic pressure
\label{sec:grava2}}
This model assumes a thick shell with continuous profile of anisotropic pressure to avoid the
introduction of an infinitesimally thin shell. The stress-energy tensor is $T^{\mu}{}_{\nu}=
\textrm{diag}[-\rho, p_r, p_t, p_t]$, where $p_r$ and $p_t$ are the radial and tangential
pressures, respectively. The density function is
\beq
\rho(r) = \left\{ \begin{array}{lll} \rho_0\,, & 0 \le r \le r_1 &\qquad \textrm{region I}\nonumber\\
ar^3+br^2+cr+d\,, & r_1 < r < r_2 &\qquad \textrm{region II}\nonumber\\
0\,, & r_2 \le r                  &\qquad \textrm{region III}
\end{array} \right.\,
\label{rho}
\eeq
with boundary conditions $\rho(0)=\rho(r_1) = \rho_0$, $\rho(r_2) = \rho'(r_1) = \rho'(r_2) =
0$ and
\beq a&=& \frac{2\rho_0}{(r_2-r_1)^3}\,, \quad b =
-\frac{3\rho_0(r_2+r_1)}{(r_2-r_1)^3}\,, \\ c &=&
\frac{6\rho_0r_1r_2}{(r_2-r_1)^3}\,, \quad d =
\frac{\rho_0(r_2^3-3r_1r_2^2)}{(r_2-r_1)^3}\,. \eeq
The density is related to the total mass $M$ by
\be \frac{\rho_0}{M}
=\frac{15}{2\pi(r_1+r_2)(2r_1^2+r_1r_2+2r_2^2)}\,.\ee
The radial pressure $p_r$ is chosen as \cite{Chirenti:2007mk}
\be p_r(\rho) = \left(\frac{\rho^2}{\rho_0}\right)
\left[\alpha-(1+\alpha)\left(\frac{\rho}{\rho_0}\right)^2\right] \,,
\label{eos}\ee
where the parameter $\alpha$ is determined by demanding that the maximum sound speed coincides
with the speed of light. (This requirement rules out superluminal behavior and implies
$\alpha\sim 2.21$.) The metric coefficients are
\be
f=\left(1-\frac{2M}{r_2}\right)
e^{\Gamma(r)-\Gamma(r_2)}\,,\quad  \frac{1}{B}=1-\frac{2m(r)}{r}\,,
\label{lambda}
\ee
where
\be
m(r)=\int_0^r 4\pi r^2\rho d r\,,
\nonumber
\ee
and
\be
\Gamma(r)= \int_0^r \frac{2m(r)+8\pi r^3 p_r}{r(r-2m(r))}d r\,.
\label{Gamma}
\ee
The above equations completely determine the structure of the gravastar. Both the metric and
its derivatives are continuous across $r_2$ and throughout the spacetime. The behaviors of the
metric coefficients for a typical gravastar are shown in the bottom panel of
Fig.~\ref{fig:metricgrava}.
\subsubsection{\label{sec:ergoregion} Slowly rotating rigid gravastars
and ergoregions}
There are no known solutions describing rotating gravastars. Thus an analysis of the ergoregion
instability for these objects is nontrivial. Fortunately, slowly rotating solutions can be
obtained using the formalism developed in Ref.~\cite{Hartle:1967he}, which we now extend to the
case of anisotropic stresses.

A rotation of order $\Omega$ gives corrections of order $\Omega^2$ in the diagonal coefficients
of the metric (\ref{metricspherical}) and introduces a non-diagonal term of order $\Omega$,
\be
g_{t\phi}\equiv -\zeta g_{\phi\phi}\,,
\ee
where $\phi$ is the azimuthal coordinate. The metric coefficient $g_{t\phi}$ defines the
angular velocity of frame dragging $\zeta=\zeta(r)$. The full metric is
\beq
ds^2&=&-f(r)dt^2+B(r) dr^2+r^2d\theta^2+\nonumber\\
&&\null\hskip36pt +r^2\sin^2\theta\left(
d\phi-\zeta(r) dt\right )^2\,.
\label{rotfull}
\eeq
We consider the anisotropic fluid stress-energy tensor
\be
T^{\mu\nu}=(\rho+p_t)U^{\mu}U^{\nu}+p_tg^{\mu\nu}+(p_r-p_t)s^\mu
s^\nu \,,
\label{Tmunu}
\ee
where
\beq
&&U^{\mu}U_{\mu}=-1\,,\quad s^{\mu}s_{\mu}=1\,,\quad U^{\mu}
s_{\mu}=0\,,\nonumber\\
&&U^r=U^{\theta}=0\,,\quad U^{\phi}=\Omega U^t\,,\nonumber \\
&&U^t=\left [-\left (g_{tt}+2\Omega
g_{t\phi}+\Omega^2g_{\phi\phi}\right )\right]^{-1/2}\,.\nonumber
\eeq
Equation (\ref{Tmunu}) describes an anisotropic fluid with radial pressure $p_r$ and tangential
pressure $p_t$, rotating with angular velocity $\Omega$ as measured by an observer at rest in
the $(t\,,r\,,\theta\,,\phi)$ coordinates. If the gravastar rotates rigidly, i.e.\ $\Omega={\rm
constant}$, the Einstein equations at order $\zeta$ give
\beq
-8\pi \rho &=& \frac{B-B^2-rB'}{r^2B^2}\,,\label{e1}\\
8\pi p_r &=&\frac{f-Bf+rf'}{r^2Bf}\,,\label{e2}\\
8\pi p_t &=& -\frac{2f^2B'+rBf'^2}{4rB^2f^2}-\nonumber\\
&&-\frac{f\left (rB'f'-2B(f'+rf'')\right
)}{4rB^2f^2} \,.\label{e3}
\eeq
An equation for $\zeta(r)$ is obtained by considering
\be
R_{t\phi}=8\pi\left (T_{t\phi}-\frac{1}{2}g_{t\phi}T\right
)\label{rtphi}\,.
\ee
Using Eqs.\ (\ref{e1}) and (\ref{e2}), Eq.\ (\ref{rtphi}) is written as
\be
\zeta''+\zeta'\left (\frac{4}{r}+\frac{j'}{j}\right ) =16\pi
B(r) (\zeta-\Omega) \left (\rho+p_t\right )\,,
\label{zeta}
\ee
where $j\equiv (fB)^{-1/2}$ is evaluated at zeroth order and $\rho$, $p_t$ are given in terms
of the nonrotating geometry by Eq.~(\ref{e1}) and Eq.~(\ref{e3}), respectively. The above
equation reduces to the corresponding equation in Ref.~\cite{Hartle:1967he} for isotropic
fluids. Solutions of Eq.~(\ref{zeta}) describe rotating gravastars to first order in $\Omega$.

Spinning gravastars may possess ergoregions. A simple but general procedure to determine their
presence for slowly rotating stars is described in Ref.\ \cite{schutzexistenceergoregion}. This
method requires only a knowledge of the metric of nonrotating objects and compares favorably
with more sophisticated numerical analyses \cite{ipserbutterworth}.

The ergoregion can be found by computing the surface on which $g_{tt}$ vanishes
\cite{schutzexistenceergoregion}:
\be
0=-f(r)+\zeta^2 r^2\sin^2\theta\,.
\label{eqergo}
\ee
Equation (\ref{eqergo}) is expected to be a good approximation to the location of the
ergoregion specially for very compact stars \cite{schutzexistenceergoregion}. The solution of
Eq.~(\ref{eqergo}) is topologically a torus. In the equatorial plane we have
\be
r\zeta(r)=\sqrt{f(r)}\,.
\label{eqergo2}
\ee
The existence and the boundaries of the ergoregions can be computed from the above equations.
Equation (\ref{zeta}) is integrated from the origin with initial conditions $(\Omega-\zeta)'=0$
and $(\Omega-\zeta)$ finite. Changing the value of $(\Omega-\zeta)$, the whole space of slowly
rotating gravastars can be obtained. The exterior solution satisfies
$\Omega-\zeta=\Omega(1-2I/r^3)$, where $I$ is the moment of inertia of the star. Demanding the
continuity of both $(\Omega-\zeta)'$ and $(\Omega-\zeta)$, $\zeta$ and $I$ are uniquely
determined. The rotation parameter $\Omega$ depends on the initial condition at the origin.

\begin{figure}[ht]
\begin{center}
\begin{tabular}{l}
\epsfig{file=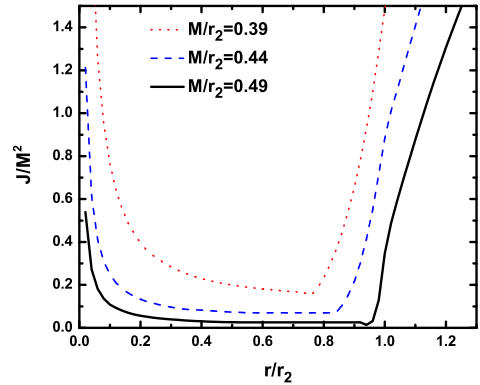,width=210pt,angle=0}\\
\null\hskip
16pt\epsfig{file=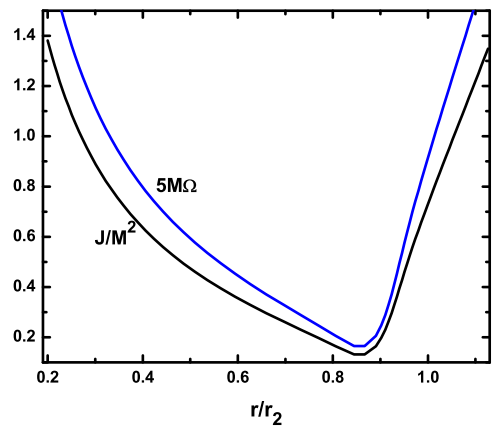,width=194pt,angle=0}
\end{tabular}
\end{center}
\caption{Top panel: Size of the ergoregion for three different gravastars in the thin-shell
limit. The vertical axis gives the angular momentum of the gravastar in units of its total
mass. The horizontal axis gives the locations of the ergoregion boundaries in units of the
gravastar radius $r_2$. Each curve refers to a different gravastar. The minima of the curves
determine the existence and extent of ergoregions. The size of an ergoregion can be found by
drawing a horizontal line at a given value of $J/M^2$; its intersections with the curves give
the radii of the ergoregion boundaries. From top to bottom the three curves refer to $r_2=1.3$,
$r_1=1$, $w_1=50$ and $w_2=1$, corresponding to $M\sim 0.39\,r_2$; $r_2=1.2$, $r_1=1$,
$w_1=150$ and $w_2=1$, corresponding to $M\sim 0.44\,r_2$; $r_2=1.05$, $r_1=1$, $w_1=350$ and
$w_2=1$, corresponding to $M\sim 0.49\,r_2$. Bottom panel: $J/M^2$ and angular frequency
$\Omega$ for the anisotropic pressure model with $r_2=2.2$, $r_1=1.8$ and $M=1$. The angular
frequency is always very small, thus the slow-rotation formalism applies. These results extend
up to the Keplerian frequency $\Omega_K$.}
\label{fig:ergoregion}
\end{figure}

Figure \ref{fig:ergoregion} shows the results for three different gravastars described in the
previous sections. The minima of the curves are the minimum values of $J/M^2$ which are
required for the existence of the ergoregion. Comparison with the results for stars of uniform
density \cite{schutzexistenceergoregion}, shows that ergoregions form more easily around
gravastars due to their higher compactness. Figure \ref{fig:ergoregion} also shows that the
ergoregions spread inside the gravastar. (The ergoregion can be located by drawing an
horizontal line at the desired value of $J/M^2$, as explained in the caption.)

Gravastars spinning above a given threshold are not stable against mass shedding
\cite{Stergioulas:2003yp}. Instability arises when the centrifugal force is strong enough to
disrupt the star. In Newtonian gravity, the equatorial mass shedding frequency is approximately
the Keplerian frequency $M\Omega_K=(M/R)^{3/2}$. Although corrections to the Keplerian
frequency are expected in a general relativistic framework, $\Omega_K$  provides a good
estimator for the validity of the slow-rotation approximation. (See Ref.~\cite{Berti:2004ny}
for a comparison of the slow-rotation regime vs.\ full numerical results.) In the following,
the slow-rotation approximation will be considered valid for $\Omega/\Omega_K < 1$. Numerical
results extend up to $\Omega \sim \Omega_K$.
\subsection{\label{subsec:bosonstars} Boson stars}
A well-known example of nonrotating boson star is the model by Colpi, Shapiro and Wasserman
(CSW) \cite{Colpi:1986ye}. A variation of the CSW model
which allows for rotating solutions is the Kleihaus, Kunz, List and
Schaffer (KKLS) model \cite{Kleihaus:2005me}. The KKLS solution is
based on the self-interacting complex scalar field $\Phi$ with Lagrangian
density
\be {\cal L}_{KKLS}=-\frac{1}{2} g^{\mu\nu}\left( \Phi_{, \, \mu}^*
\Phi_{, \, \nu} + \Phi _ {, \, \nu}^* \Phi _{, \, \mu} \right) - U(
\left| \Phi \right|)\,, \label{KKL} \ee
where $U(|\Phi|)=\lambda |\Phi|^2(|\Phi|^4-a|\Phi|^2+b)$. The mass of the boson is given by
$m_{\rm B}=\sqrt{\lambda b}$. The equations for the boson star structure can be solved by
setting
\beq
ds^2 =- f dt^2 &+& \frac{k}{f} \, \biggl[ g \left( dr^2 + r^2 \, d\theta^2 \right)\nonumber\\
 &+& r^2 \, \sin^2 \theta \,  \, \left( d \varphi
-\zeta(r) \, dt \right)^2 \biggr]
\label{ansatzg}
\eeq
and $\Phi=\phi~e^{ i\omega_s t +i n \varphi}$, where the metric components and the real
function $\phi$ depend only on $r$ and $\theta$. The requirement that $\Phi$ is single-valued
implies $n=0,\pm 1,\pm 2,\dots$. The solution has spherical symmetry for $n=0$ and axial
symmetry otherwise. The mass $M$ and the angular momentum $J$ can be read off from the
asymptotic expansion of $f$ and $\zeta$,
\be M=\frac{1}{2G} \lim_{r \rightarrow \infty} r^2\partial_r \, f \
, \ \ \  J=\frac{1}{2G} \lim _{r \rightarrow \infty} r^3\zeta\,,
\label{MJ2} \ee
respectively. Since the Lagrangian density is invariant under a global $U(1)$ transformation,
the current $j^{\mu}=-i\Phi^*\partial^{\mu}\Phi+{\rm c.c.}$ is conserved. The associated charge
is
\be Q =4 \pi \omega_s \int_0^{\infty}\int _0^{\pi}
|g|^{1/2}\frac{1}{f} \left(1 + \frac{n}{\omega_s}\frac{\omega}{r}
\right) \phi^2 \, dr \, d\theta \,.
\label{Qc}
\ee
\begin{center}
\begin{figure*}[ht]
\begin{tabular}{cc}
\epsfig{file=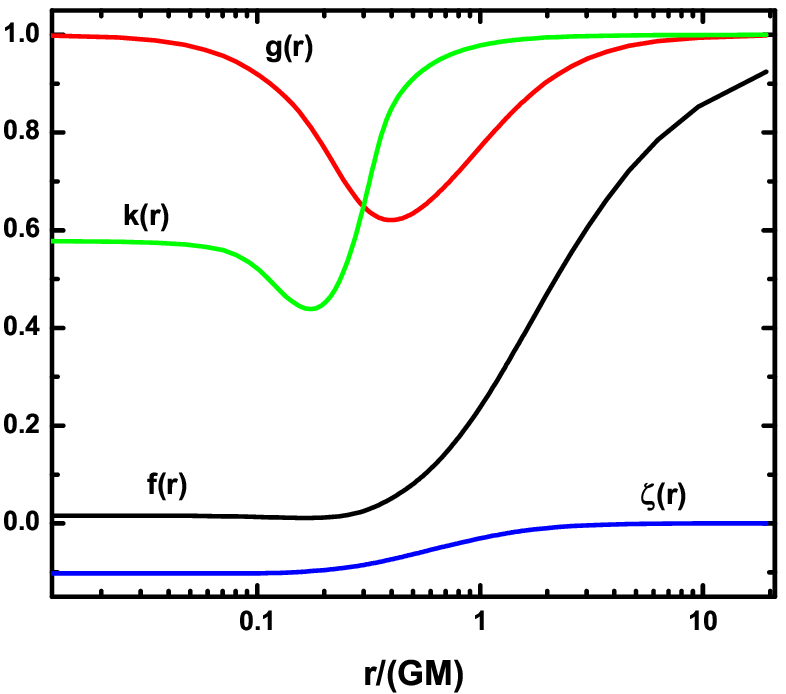,width=210pt,angle=0}
\epsfig{file=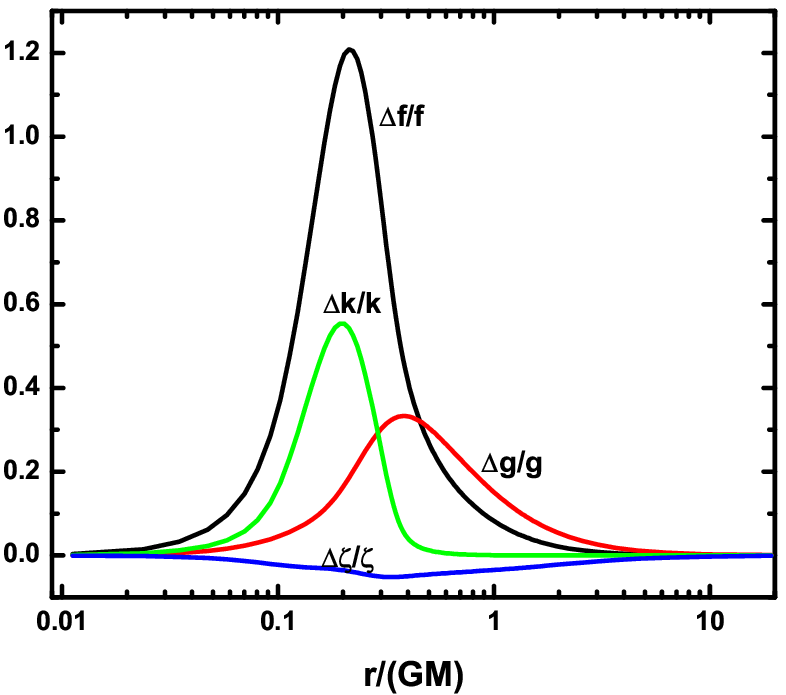,width=210pt,angle=0}&
\end{tabular}
\caption{Left panel: Metric coefficients for a rotating boson star along the equatorial
plane, with parameters $n=2$, $b=1.1$, $\lambda=1.0$, $a=2.0$, $J/(GM^2)\sim 0.566$.  Right
panel: Fractional difference of the metric potentials between $\theta=\pi/2$ and
$\theta=\pi/4$ for the same star. The plot gives the maximum possible fractional difference
between these quantities.}
\label{fig:metricrotboson}
\end{figure*}
\end{center}
It can be shown that the angular momentum $J$ and the scalar
charge $Q$ of stationary solutions, i.e., solutions with a
timelike and rotational Killing vector, are related by the
quantization condition $J=nQ$ \cite{schunck}. This may erroneously
lead to the conclusion that boson stars are gravitationally stable
because a continuous extraction of angular momentum is needed to
trigger instability. However, the above quantization condition
applies for objects for which such Killing vectors can be defined
and we will assume that this condition is broken in the presence
of perturbations. In this case, an arbitrary amount of angular
momentum can be extracted and it will be shown below that
gravitational-wave emission leads to an instability on very short
timescales.

The numerical procedure to extract the metric and the scalar field is described in Ref.\
\cite{Kleihaus:2005me}. Throughout the paper we will consider solutions with $n=2$, $b=1.1$,
$\lambda=1.0$, $a=2.0$ and different values of $(J\,,M)=(3781,1296), (3400,1081), (2800,906)$,
corresponding to $J/(GM^2)\sim 0.566$, $0.731$ and $0.858$, respectively. The  $n=1$ solutions
in Ref.~\cite{Kleihaus:2005me} exhibit similar features. The two top panels of
Fig.~\ref{fig:metricrotboson} show the metric functions for boson stars with $J/(GM^2)\sim
0.566$ and $0.858$ along the equatorial plane. The change in the metric potentials from
$\theta=\pi/2$ to $\theta=\pi/4$ for these solutions is plotted in the bottom panels of
Fig.~\ref{fig:metricrotboson}. The metric functions do not depend significantly on the
longitudinal angle.
\begin{figure}
\epsfig{file=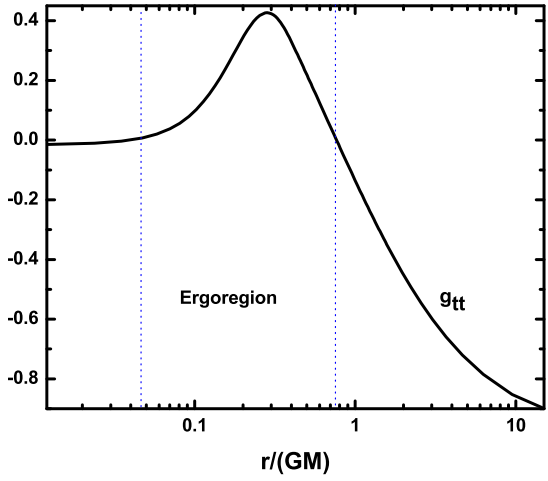, width=.45\textwidth,
angle=0} \caption{The $g_{tt}$ metric coefficient for a boson star
with  $J/(GM^2)\sim 0.566$ at its equator. The ergoregion is
identified by the region inside the dotted vertical lines and
extends from $r/(GM)\sim 0.047$ to $0.770$.}
\label{fig:metricrotbosongtt}
\end{figure}
Figure \ref{fig:metricrotbosongtt} gives $g_{tt}$ as a function of distance for the case with
$J/(GM^2)\sim 0.566$ at the equator.  The behavior of $g_{tt}$ demonstates that boson stars
develop ergoregions deeply inside the star. For this particular choice of parameters, the
ergoregion extends from $r/(GM)\sim 0.0471$ to $0.770$. A more complete discussion on the
ergoregions of rotating boson stars can be found in Ref.~\cite{Kleihaus:2005me}.
\section{Ergoregion instability for rotating stars}
\label{sec:ergoinst}
The stability of ultra-compact objects can be studied perturbatively by considering small
deviations around equilibrium. As explained in the introduction, we consider only scalar
perturbations. This is justified as follows. Axial gravitational perturbations are described in
the large $l=m$ limit by the same equation of scalar perturbations. In this regime, our results
describe both kinds of perturbations. In the small $l=m$ limit, gravitational perturbations are
expected to have shorter growth times than scalar perturbations: black holes are characterized
by a superradiant amplification of spin-2 fields which is much stronger than the superradiant
amplification of other fields. This is due to the potential barrier outside the horizon having
different behavior for different spin-field $s$. Since the ultra-compact objects we are dealing
with are also characterized by a relativistic potential barrier, gravitational perturbations
are expected to couple more strongly to the ergoregion and have smaller instability timescales.
This conclusion is also verified under certain simplifying assumptions in
Ref.~\cite{Kokkotas:2002sf}. Thus, scalar perturbations should provide a lower bound on the
strength of the instability.
\subsection{Axial gravitational perturbations for perfect fluid stars}
\label{sec:ergograv}
In the large $l=m$ regime, axial gravitational perturbations \cite{Ruoff:2001wr,
Kokkotas:2002sf} are described by a simple equation. In general, axial and polar
perturbations are coupled when rotation is included \cite{Kojima:1992ie}. For simplicity, we
will assume that the zeroth-order polar perturbations vanish and there is no coupling. The
full metric is a perturbation of Eq.~(\ref{rotfull}) \cite{Kojima:1992ie}:
\beq
ds^2&=&ds_0^2+2\sum _{lm}\left
(h_0^{lm}(t,r)+h_1^{lm}(t,r)\right ) \nonumber \\
&&\times \left (-\sin^{-1}\theta
\partial_{\phi}Y_{lm}d\theta+\sin\theta\partial_{\theta}Y_{lm}d\phi\right )\,,
\eeq
where $ds_0^2$ is the unperturbed metric (\ref{rotfull}) and $Y_{lm}$ are scalar spherical
harmonics. The quantities
\be h_0^{lm}\equiv\sqrt{f(r)/B(r)}K_6\,,\qquad h_1^{lm}\equiv
\sqrt{B(r)/f(r)}\,V_4\,, \ee
satisfy the system of equations (see Eqs.~(13)-(16) in
Ref~\cite{Ruoff:2001wr})
\be
K_3'=16\pi(p+\rho)u_3-\frac{2K_3}{r}+\frac{l^2+l-2}{r^2}K_6-\frac{2m\zeta'V_4}
{(l^2+l)f}\,,
\label{axgrav1}
\ee
\be
K_6'=-\frac{B}{f}(-\omega+m\zeta)V_4-\left
(\frac{f'}{2f}-\frac{B'}{2B}-\frac{2}{r}\right )K_6+BK_3\,,
\label{axgrav2}
\ee
where $K_3$ and $K_6$ are two extrinsic curvature variables and
\beq
&&\null\hskip-24pt
V_4=\frac{r^2}{l^2+l-2}\left
((-\omega+m\zeta)K_3-\frac{2m\zeta'}{l(l+1)}\frac{K_6}{B}\right)\,,\\
&&\null\hskip-24pt
u_3=\frac{2m
(\Omega-\zeta)}{2m(\Omega-\zeta)-l(l+1)(-\omega+m\Omega)}K_6\,.
\eeq
Here and throughout this paper, it is understood that
$\omega\equiv \omega_{lm}$. In the large $l=m$ limit,
Eqs.~(\ref{axgrav1}) and (\ref{axgrav2}) reduce to
\beq
K_3'&=&-\frac{2}{r}K_3+\frac{m^2}{r^2}K_6\,,\label{axgrav12}\\
K_6'&=&BK_3-\frac{B}{f}(\Sigma+\zeta)^2r^2K_3-\left
(\frac{f'}{2f}-\frac{B'}{2B}-\frac{2}{r}\right )K_6\,,\nonumber\\
\label{axgrav22}
\eeq
where $\Sigma\equiv -\omega/m$. Combining Eqs.~(\ref{axgrav12})-(\ref{axgrav22}) and neglecting
terms of order $1/m^2$, it follows
\be
K_3''+m^2\frac{B}{f}\left ((\Sigma+\zeta)^2-\frac{f}{r^2}\right
)K_3=0\,.
\label{axgravfinal}
\ee
This equation also describes the scalar perturbations of gravastars, as it will be shown below.
\begin{figure}
\begin{center}
\begin{tabular}{c}
\epsfig{file=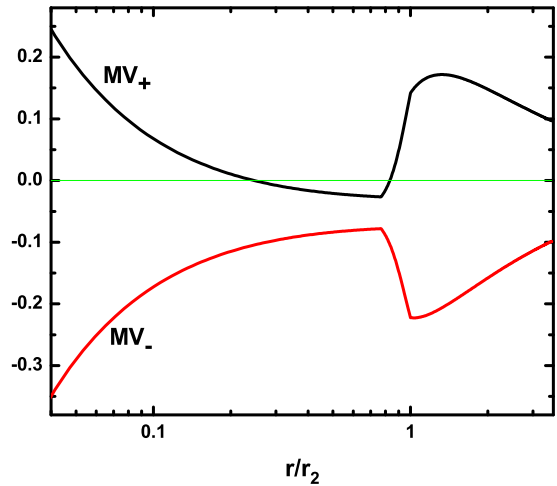,width=220pt,angle=0}\\
\epsfig{file=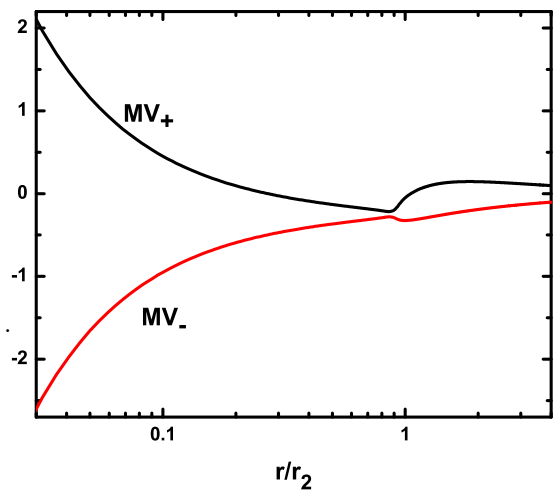,width=220pt,angle=0}
\end{tabular}
\end{center}
\caption{Top panel: Potentials $V_{\pm}$ for the thin-shell
gravastar with $r_2=1.3$, $r_1=1$, $w_1=50$ and $w_2=1$. The
ergoregion extends from $r\sim 0.247r_2$ to $0.832r_2$ and
corresponds to a gravastar with $J\sim 0.333M^2$ and $M\Omega\sim
0.105$. Bottom panel: Potentials for the anisotropic pressure
gravastar with $r_2=2.2$, $r_1=1.8$ and $M=1$. The ergoregion
extends from $r\sim 0.270r_2$ to $r\sim 1.055r_2$ and rotates with
angular momentum $J/M^2=1.00$, corresponding to $\Omega\sim 0.250$.}
\label{fig:poyergoregioninstgrava}
\end{figure}

\subsection{\label{sec:ergoregioninstslowlyrot} Scalar field instability for slowly rotating gravastars: WKB
approach}
Consider now a minimally coupled scalar field in the background of a gravastar. The metric of
gravastars is given by Eq.~(\ref{rotfull}). In the large $l=m$ limit, which is appropriate for
a WKB analysis \cite{cominsschutz, Cardoso:2005gj}, the scalar field can be expanded as
\be \Phi=\sum_{lm}\bar{\chi}_{lm}(r)\exp{\left [-\frac{1}{2}\int
\left (\frac{2}{r}+\frac{f'}{2f}+\frac{B'}{2B}\right )dr
\right]}e^{-i\omega t}Y_{l
m}(\theta\,,\phi)\,.\label{scalardecomposition} \ee
The functions $\bar{\chi}_{lm}$ are determined by the Klein-Gordon
equation, which yields
\be \bar{\chi}_{lm}''+m^2T(r\,,\Sigma)\bar{\chi}_{lm}=0\,,
\label{equationphigrava} \ee
where $\Sigma$ is defined as below Eq.~(\ref{axgrav22}) and
\beq
T&=&\frac{B(r)}{f(r)}\left (\Sigma-V_+\right )\left (\Sigma-V_-\right )\,,\\
V_{\pm}&=&-\zeta \pm \frac{\sqrt{f(r)}}{r}\,.
\label{TWKB}
\eeq
Equation (\ref{equationphigrava}) follows from Eq.~(\ref{scalardecomposition}) when terms of
order ${\cal O}\left (1/m^2 \right)$ are dropped. Equation (\ref{equationphigrava}) can be
shown to be identical to Eq.~(\ref{axgravfinal}) for the axial gravitational perturbations of
perfect fluid stars. Therefore, the following results apply to both kinds of perturbations.

The eigenfrequencies of Eq.~(\ref{equationphigrava}) can be computed in the WKB approach
following Ref.~\cite{cominsschutz}. This method is in excellent agreement with full numerical
results \cite{eriguchi,Cardoso:2005gj}. The quasi-bound unstable modes are determined by
\be
m\int_{r_a}^{r_b}\sqrt{T(r)}dr =\frac{\pi}{2}+n\pi\,,\quad
n=0\,,1\,,2\,,\dots
\ee
and have timescale
\be
\tau=4\exp{\left[2m\int_{r_b}^{r_c}\sqrt{|T|}dr\right]}\int_{r_a}^{r_b}\frac{d}{d\Sigma}
\sqrt{T}dr\,,
\ee
where $r_a$, $r_b$ are solutions of $V_+=\Sigma$ and $r_c$ is determined by the condition
$V_-=\Sigma$.

The potentials $V_{\pm}$ are displayed in Fig.~\ref{fig:poyergoregioninstgrava} for the
gravastar models of Sect.~\ref{subsec:gravastar}. The top panel shows the potential for the
thin-shell model with $r_2=1.3$, $r_1=1$, $w_1=50$ and $w_2=1$. The gravastar rotates with
angular frequency $\Omega\sim 0.105$ and the ergoregion lies in the region
$r\sim(0.247,0.832)r_2$. The bottom panel refers to the anisotropic pressure model with
$r_2=2.2$, $r_1=1.8$ and $M=1$. The ergoregion extends from $r\sim 0.270\, r_2$ to $r\sim
1.055\, r_2$ and rotates with angular frequency $\Omega\sim 0.250$.

\begin{figure}[ht]
\begin{tabular}{c}
\epsfig{file=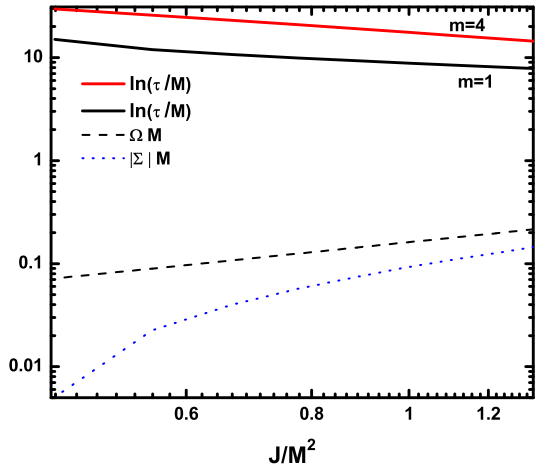,width=240pt,angle=0}
\end{tabular}
\caption{Details of the ergoregion instability ($m=1$ and $m=4$) for the thin-shell gravastar
of Sect.\ \ref{sec:grava1} with $r_2=1.3$, $r_1=1$, $w_1=50$ and $w_2=1$. The plot shows the
logarithm of the dimensionless instability timescale $\tau/M$, the dimensionless angular
velocity $M\Omega$ and the oscillation frequency $\Sigma$ vs.\ the angular momentum per unit
mass, $J/M^2$. The $m=2,3$ modes follow a similar pattern and are in between the $m=1,4$ results.}
\label{fig:ergograva}
\end{figure}

The results of the WKB computation are shown in Fig.~\ref{fig:ergograva} and Tables
\ref{tab:instrotgrava}-\ref{tab:instrotgravaII}. Figure \ref{fig:ergograva} displays the
results for the least compact thin-shell gravastar of Sect.~\ref{subsec:gravastar} with
$m=1$, $4$. Although the WKB approximation breaks down at low $m$ values, these results
still provide reliable estimates \cite{cominsschutz}. This claim will be verified in
Sect.~\ref{sec:ergoregioninstnumerical} with a full numerical integration of the
Klein-Gordon equation. Table \ref{tab:instrotgrava} compares three different gravastars for
$J/M^2=1$ and $m=1\,,2\,,\dots 5$. The results show that the instability timescale decreases
as the star becomes more compact. Table \ref{tab:instrotgravaI} refers to the most compact
thin-shell gravastar for various angular frequencies. The instability timescale depends
strongly on the rotation. A fit for the instability timescale in powers of $J/M^2$ yields
\be
\log \tau/M \sim a+b\sqrt{J/M^2}+cJ/M^2\,,
\ee
where $a=68.0$, $b=-76.7$, $c=26.2$ for $m=5$ and $a=55.8$, $b= -61.4$, $c=21.0$ for $m=4$,
respectively. The results for lower $m$ values show similar behaviors. For instance, for $m=1$
the coefficients are $a=19.7, b=-16.2$ and $c=5.6$. Table \ref{tab:instrotgravaII} shows the
WKB results for the anisotropic pressure model for different values of $J/M^2$. Larger values
of $J/M^2$ make the star more unstable. The instability timescales are fitted by
\be
\log \tau/M \sim a+b\left (J/M^2\right )^c\,,
\ee
where $a=-21.9$, $b=39.2$, $c=-0.39$ for $m=5$ and $a=-13.7$, $b=28.5$, $c= -0.43$ for $m=4$.
The trend for lower-$m$ is similar. The coefficients for $m=1$ are $a=7.4, b=0.57, c=-3.0$.

Both models have similar low-$m$ behaviors. It will be shown in
Sect.~\ref{sec:ergoregioninstnumerical} that the WKB results for the instability timescale
differ from the numerical results by about one order of magnitude at low $m$. On the contrary,
the resonant frequencies match well the WKB results even for low-$m$ modes. Calculations show
that the resonant frequency is ${\rm Re(\omega)}\sim \alpha \Omega$, where $\alpha\sim 1.1$ --
$1.2$.

The maximum growth time of the instability is of the order of a few thousand $M$, at least
for large $J$. This instability is crucial for the star evolution. Comparison of Table
\ref{tab:instrotboson} with Tables 1 and 2 of Ref.~\cite{cominsschutz} shows that the
ergoregion instability of gravastars is stronger than the ergoregion instability of uniform
density stars by many orders of magnitude. This seems to be a general feature of all
ultra-compact objects discussed here. Gravitational perturbations are expected to be even
more unstable.

\begin{table}
\centering
\caption{\label{tab:instrotgrava} WKB results for the instability of rotating thin-shell
gravastars with $J/M^2=1$. The instability grows with compactness.}
\begin{tabular}{|c||ccc|}
%
\hline \multicolumn{1}{|c}{} & \multicolumn{3}{c|}{ $(-10^2M \Sigma\,,\tau/M)$}\\
\hline
$m$  & $r_2=1.3$&$r_2=1.2$&$r_2=1.05$    \\
1    & $9.2\,,6.85\times10^3$ &$17\,,5.26\times10^3$&$23\,,9.16\times10^3$\\
2    & $11\,,1.23\times10^5$ &$18\,,5.63\times10^4$&$23\,,7.37\times10^4$\\
3    & $12\,,2.31\times10^6$ &$18,,6.06\times10^5$&$23\,,5.98\times10^5$\\
4    & $12\,,4.34\times10^7$ &$18\,,6.58\times10^6$&$23\,,4.86\times10^{6}$\\
5    & $12\,,8.26\times10^8$ &$18\,,7.13\times10^7$&$23\,,3.99\times10^{7}$\\
\hline \hline
\end{tabular}
\end{table}
%

\subsubsection{Comparison with numerical results}
\label{sec:ergoregioninstnumerical}
Accurate computations of the instability require numerical solutions of the Klein-Gordon
equation. However, the WKB approximation provides reliable estimates of the numerical
results. The exact potential $T$ of Eq.~(\ref{TWKB}) is
\beq \bar{T}&=&
-\frac{l(l+1)B}{r^2}+\frac{B\left(\omega-m\zeta\right
)^2}{f}+\frac{B'}{2rB}+\frac{B''}{4B}+\nonumber
\\
&&-\frac{5B'^2}{16B^2}-\frac{f'}{2rf}+\frac{B'f'}{8Bf}
+\frac{3f'^2}{16f^2}-\frac{f''}{4f}\,. \eeq
The results of the numerical integration for the anisotropic pressure gravastar with $r_2=2.2$,
$r_1=1.8$, $M=1$ and $J/M^2=1$ are shown in Table \ref{tab:gravcomp}.

\begin{widetext}
\begin{center}
\begin{table}[ht]
\centering
\caption{\label{tab:instrotgravaI} WKB results for the instability of rotating thin-shell
gravastars with $r_2=1.05$ and $r_1=1.0$. The mass shedding limit corresponds to
$\Omega/\Omega_K=1$, where $\Omega_K=\sqrt{M/r_2^3}$.}
\begin{tabular}{|c||ccccccc|}
%
\hline \multicolumn{1}{|c}{} & \multicolumn{7}{c|}{ $(-10^2M \Sigma\,,\tau/M)$}\\
\hline
&$J/M^2=0.40$    &$J/M^2=0.50$     &$J/M^2=0.60$    &$J/M^2=0.70$   &$J/M^2=0.80$    &$J/M^2=0.90$     &$J/M^2=1.0$\\
$m$&$\Omega/\Omega_K=0.28$&$\Omega/\Omega_K=0.35$&$\Omega/\Omega_K=0.42$&$\Omega/\Omega_K=0.50$&$\Omega/\Omega_K=0.57$&$\Omega/\Omega_K=0.64$&$\Omega/\Omega_K=0.71$\\
1&$8.4,1.19\t10^{5}$&$10,6.28\t10^{4}$&$13,3.70\t10^{4}$&$16,2.38\t10^{4}$&$18,1.64\t10^{4}$&$20,1.19\t10^{4}$&$23,9.15\t10^{3}$\\
2&$8.7,1.13\t10^{7}$ &$11,3.24\t10^{6}$&$14,1.15\t10^{6}$&$16,4.82\t10^{5}$&$18,2.31\t10^{5}$&$21,1.24\t10^{5}$&$23,7.38\t10^{4}$\\
3&$8.8,1.08\t10^{9}$&$11,1.68\t10^{8}$&$14,3.59\t10^{7}$&$16,9.86\t10^{6}$&$18,3.30\t10^{6}$&$21,1.30\t10^{6}$&$23,5.97\t10^{5}$\\
4&$8.9,1.04\t10^{11}$&$11,8.75\t10^{9}$&$14,1.13\t10^{9}$&$16,2.03\t10^{8}$&$18,4.69\t10^{7}$&$21,1.37\t10^{7}$&$23,4.86\t10^{6}$\\
5&$8.9,1.01\t10^{13}$&$11,4.58\t10^{11}$&$14,3.59\t10^{10}$&$16,4.16\t10^{9}$&$18,6.75\t10^{8}$&$21,1.45\t10^{8}$&$23,3.98\t10^{7}$\\
\hline \hline
\end{tabular}
\end{table}
\end{center}
%
\begin{center}
\begin{table}[ht]
\centering
\caption{\label{tab:instrotgravaII} WKB results for the instability of rotating anisotropic
pressure gravastars with $r_2=2.2$, $r_1=1.8$ and $M=1$.}
\begin{tabular}{|c||ccccccc|}
%
\hline \multicolumn{1}{|c}{} & \multicolumn{7}{c|}{ $(-10^2M \Sigma\,,\tau/M)$}\\
\hline
&$J/M^2=0.40$    &$J/M^2=0.50$     &$J/M^2=0.60$    &$J/M^2=0.70$   &$J/M^2=0.80$    &$J/M^2=0.90$     &$J/M^2=1.0$\\
$m$&$\Omega/\Omega_K=0.33$&$\Omega/\Omega_K=0.41$&$\Omega/\Omega_K=0.49$&$\Omega/\Omega_K=0.57$&$\Omega/\Omega_K=0.65$&$\Omega/\Omega_K=0.74$&$\Omega/\Omega_K=0.82$\\
1&$0.24,1.33\t10^{7}$&$2.7,1.13\t10^{5}$&$5.2,2.78\t10^{4}$&$7.6,1.15\t10^{4}$&$10,5.99\t10^{3}$&$12,3.58\t10^{3}$&$15,2.34\t10^{3}$\\
2&$3.1,8.25\t10^{7}$ &$5.6,6.20\t10^{6}$&$8.1,1.14\t10^{6}$&$10,3.13\t10^{5}$&$13,1.11\t10^{5}$&$15,4.81\t10^{4}$&$18,2.33\t10^{4}$\\
3&$4.2,1.31\t10^{10}$&$6.6,5.44\t10^{8}$&$9.1,5.65\t10^{7}$&$12,9.40\t10^{6}$&$14,2.25\t10^{6}$&$17,6.82\t10^{5}$&$19,2.45\t10^{5}$\\
4&$4.7,2.50\t10^{12}$&$7.2,5.13\t10^{10}$&$9.7,2.95\t10^{9}$&$12,3.10\t10^{8}$&$15,4.81\t10^{7}$&$17,1.02\t10^{7}$&$20,2.73\t10^{6}$\\
5&$5.1,5.06\t10^{14}$&$7.6,4.99\t10^{12}$&$10,1.59\t10^{11}$&$13,9.82\t10^{9}$&$15,1.02\t10^{9}$&$17,1.52\t10^{8}$&$20,3.07\t10^{7}$\\
\hline \hline
\end{tabular}
\end{table}
\end{center}
\end{widetext}

\begin{table}[ht]
\centering
\caption{\label{tab:gravcomp} Comparison between analytical and numerical results for
anisotropic pressure gravastars with $J/M^2=1$, $r_2=2.2$, $r_1=1.8$ and $M=1$. The numerical
results for the real part are in good agreement with the WKB results. The agreement is better
for larger values of $m$. The imaginary parts agree within an order of magnitude.}
\begin{tabular}{|c||ccc|}
%
\hline \multicolumn{1}{|c}{} & \multicolumn{3}{c|}{ $(-10^2\Sigma M\,,\tau/M)$}\\
\hline
$m$  & Analytical (A) & Numerical (N)  & $\vert{\frac{\Sigma_A-\Sigma_N}{\Sigma_A}}\vert$   \\
1    & $15.0,2.34\times 10^{3}$& $21.9\,,2.47\times 10^{3}$&31.5\%\\
2    & $18.0,2.34\times 10^{4}$& $18.6\,,1.81\times 10^{5}$&3.2\%\\
3    & $19.0,2.49\times 10^{5}$& $18.7\,,2.53\times 10^{6}$&1.6\%\\
4    & $19.6,2.74\times 10^{6}$& $19.0\,,3.33\times 10^{7}$&3.2\%\\
5    & $19.8,3.07\times 10^{7}$& $19.3\,,3.76\times 10^{8}$&2.6\%\\
\hline \hline
\end{tabular}
\end{table}
The WKB approximation for the real part of the frequency shows a remarkably good agreement with
the numerical results even at low values of $m$. For any $m>1$ this agreement is better than
5\%. The instability timescale seems to be more sensitive to the details of the WKB
integration, with a level of agreement similar to that reported in Ref.~\cite{eriguchi}. The
above results show that the WKB approximation correctly estimates the instability timescale for
all values of $m$ within an order of magnitude.
\subsection{Scalar field instability for rotating boson stars: WKB approach}
\label{sec:ergoregioninstboson}
Consider a scalar field $\Psi$ minimally coupled to the rotating boson star geometry (not to be
confused with the scalar field which makes up the star). Since the metric coefficients depend
on $r$ and $\theta$, the Klein-Gordon equation cannot be reduced, in general, to a
one-dimensional problem. Separation of variables can be achieved by requiring $g=1$,
$f=f(r,\pi/2)$, $k=k(r,\pi/2)$ and $\zeta=\zeta(r,\pi/2)$. These assumptions are justified as
follows. Firstly, the metric function $g$ is very close to unity throughout the entire
coordinate region, as can be seen in Figure \ref{fig:metricrotboson}. Since the Klein-Gordon
equation does not depend on the derivatives of $g$, it seems safe to set $g=1$ in the whole
domain. Secondly, the angular dependence of the metric coefficients is negligible for slow
rotations. The largest variation for one revolution around the star is that of  $f$, which is
less than 100\% for most cases (see Fig.~\ref{fig:metricrotboson}). Moreover, this dependence
is extremely weak for most of the values of $r$. Thirdly, perturbations are localized around
the equator in the large $l=m$ behavior. Therefore, evaluating the metric coefficients at the
equator may provide a reasonable approximation to the problem. The results obtained in this
section are expected to give an order of magnitude estimate for the instability.

\begin{figure}[ht]
\begin{center}
\begin{tabular}{c}
\epsfig{file=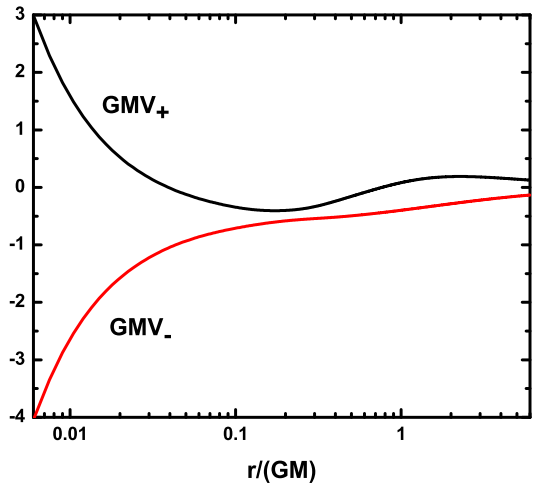,width=240pt,angle=0}
\end{tabular}
\end{center}
\caption{Potentials $V_{\pm}$ for the boson star model with $J/(GM^2)=0.566$. The ergoregion
extends from $r/(GM)\sim 0.0478$ to $0.779$.}
\label{fig:poyergoregioninstboson}
\end{figure}

The equation for the scalar field is obtained by expanding the latter as
\be \Psi=\sum_{lm} \bar{\Psi}_{lm}(r)\exp{\left[-\frac{1}{2}\int
\left (\frac{2}{r}+\frac{k'}{k}\right )dr\right]}e^{-i\omega
t}Y_{l m}(\theta\,,\phi)\,. \ee
In the large $l=m$ limit, $\bar{\Psi}_{lm}$ is determined by
\be \bar{\Psi}_{lm}''+m^2T(r\,,\Sigma)\bar{\Psi}_{lm}=0\,, \ee
where
\beq
T&=&\frac{k}{f^2}\left (\Sigma-V_+\right )\left (\Sigma-V_-\right )\,,\\
V_{\pm}&=&\zeta \pm \frac{f}{r\sqrt{k}}\,,\quad \Sigma\equiv
\frac{\omega}{m}\,,
\eeq
and terms of order ${\cal O}\left (1/m^2\right )$ have been neglected. The potentials $V_{\pm}$
are plotted in Fig.~\ref{fig:poyergoregioninstboson} for the rotating boson star with
$J/(GM^2)=0.566$. The results of the WKB computation are summarized in Table
\ref{tab:instrotboson} for $m=1\,,2\,,\dots 5$.
\begin{table}[ht]
\centering
\caption{\label{tab:instrotboson} Instability for rotating boson stars with parameters $n=2$,
$b=1.1$, $\lambda=1.0$, $a=2.0$ and different values of $J$: $J/(GM^2)=0.566$, $0.731$ and
$0.858$. The Newton constant is defined as $4\pi G=0.05$.}
\begin{tabular}{|c||ccc|}
%
\hline \multicolumn{1}{|c}{} & \multicolumn{3}{c|}{ $-10^2GM \Sigma\,,\tau/(GM)$}\\
\hline
$m$  & $\frac{J}{GM^2}=0.566139$&$\frac{J}{GM^2}=0.730677$ &$\frac{J}{GM^2}=0.857658$    \\
1    & $31\,,8.847\times 10^2$  &$6.6\,,6.303\times10^3$ &$-$\\
2    & $36\,,7.057\times10^3$   &$13\,,5.839\times10^4$ &$0.68\,,1.478\times10^6$\\
3    & $37\,,6.274\times10^4$   &$16\,,9.274.\times10^5$&$3.4\,,2.815\times10^8$\\
4    & $38\,,5.824\times10^5$   &$17\,,1.603\times10^7$ &$4.9\,,2.815\times10^{10}$\\
5    & $38\,,5.554\times10^6$   &$18\,,2.915\times10^8$ &$5.7\,,1.717\times10^{12}$\\
\hline \hline
\end{tabular}
\end{table}
An interesting feature is that the instability timescale increases with the star angular
momentum. Due to lack of sufficient numerical results we do not yet know whether this
behavior holds for all values of the angular momentum or the trend changes at some point.
The maximum growth time for this boson star model is of the order of $10^6 M$ for
$\frac{J}{GM^2}=0.857658$. This is a relatively short instability timescale, corresponding
to about one second for a one solar mass boson star.
\section{Detectability by Earth- and space-based gravitational wave
detectors\label{sec:grwaves}}
The ergoregion instability may be of interest for gravitational wave astronomy. Contrary to the
$r$-mode instability of neutron stars \cite{Owen:1998xg,Arras:2002dw}, the ergoregion
instability is not limited to solar mass objects. Thus chances of detection are larger because
the signal can fall in frequency bands where the detectors are more sensitive.
\subsection{Signal-to-noise ratio}
Detectability depends only on the energy released and the detector frequency bandwidth. The
sky-averaged signal-to-noise ratio (SNR) is \cite{Flanagan:1997sx}
\be
\rho^2=\frac{2}{5\pi^2D^2}\int df
\frac{1}{f^2S_h(f)}\frac{dE}{df}\,,
\label{snr1}
\ee
where $D$ is the distance to the source and $S_h(f)$ is the noise power spectral density of
the detector. Using $dE=2\pi fdJ/m$, the SNR for the $l=m=2$ mode, i.e.\ the mode for which
the instability is expected to be stronger, is
\be
\rho^2=\frac{2}{5\pi D^2}\int df \frac{1}{f S_h(f)}\frac{dJ}{df}\,.
\label{snr2}
\ee
Equation (\ref{snr2}) agrees with results in the literature
\cite{Owen:1998xg,Owen:2001dz,Arras:2002dw} and is independent of
the perturbation amplitude at lowest order. Higher-order
corrections, however, would contain a dependence of the SNR ratio
on the amplitude evolution. Fitting the resonant frequencies to
\be
{\rm Re}[\omega]=2\pi f \approx \alpha \Omega\,,
\label{fitfr}
\ee
one finds $\alpha\sim 1.1-1.2$. From Eq.~(\ref{fitfr}) it follows $dJ/df\approx 2\pi
I/\alpha $. Assuming the moment of inertia $I\sim 2\beta M^3$ to be roughly independent of
the angular velocity \cite{Owen:1998xg} (computations show that $\beta\sim 1$ for gravastars
to a very good approximation) Eq.~(\ref{snr2}) can be rewritten as
\cite{Owen:1998xg,Owen:2001dz,Arras:2002dw}
\be \rho^2=\frac{8\beta M^3}{5 \alpha D^2}\int_{f_{\rm min}}^{f_{\rm
max}} df \frac{1}{f S_h(f)}\,. \ee
\begin{center}
\begin{table}[ht]
\centering
\caption{\label{tab:detectability} SNR for ergoregion instability of an object at distance of
$20 {\rm Mpc}$ for LIGO/Advanced LIGO and LISA. (See text for details.)}
\begin{tabular*}{180pt}{@{\extracolsep{\fill}} |c||cccccc|}
%
\hline \multicolumn{1}{|c}{} & \multicolumn{6}{c|}{$\rho$}\\
\hline
$M/M_{\odot}$ &$10$     &$20$    &$30$   &$40$    &$50$     &$100$\\
{\rm LIGO}&$1$&$5$&$11$&$18$&$24$&$23$\\
{\rm Adv.\,LIGO} &$9$&$84$&$193$&$249$&$304$&$468$\\
\hline
\end{tabular*}
\\~\\~\\
\begin{tabular*}{180pt}{@{\extracolsep{\fill}} |c||ccccc|}
\hline \multicolumn{1}{|c}{} & \multicolumn{5}{c|}{$\rho$}\\
\hline
$M/M_{\odot}$&$10^4$    &$10^5$     &$10^6$    &$10^7$   &$10^8$   \\
{\rm LISA}&$0.07$&$21$&$2428$&$1469$&$1417$\\
\hline
\end{tabular*}
\end{table}
\end{center}
The minimum and maximum frequencies in the above integral are
chosen as $f_{\rm min}=0.9 f_{\rm max}$ (we here implicitly assume
that the amplitude of the perturbation is sufficiently large so
that $f_{\rm min}$ is reached within the typical operation time of
a detector such as LIGO or LISA, ie $\sim ~$ 1 year). This is a
conservative estimate based on a simple model for the evolution of
the system. SNRs for objects at $20 {\rm Mpc}$ distance for
LIGO/Advanced LIGO and LISA are shown in Table
\ref{tab:detectability} for $\Omega M=0.2$. Solar-mass objects are
difficult to detect, although LIGO (Advanced LIGO) could be able
to detect objects with $M\gtrsim 30M_{\odot}$ $(M\gtrsim
10M_{\odot})$. SNRs of several thousands are easily achieved for
supermassive objects. These objects could be easily observed by
LISA.
\subsection{Waveforms}
The expression for the SNR derived above is the optimal SNR. Search and detection
techniques, i.e.\ matched filtering, usually require accurate theoretical templates. The
derivation of accurate waveforms for the ergoregion instability is beyond the purpose of
this paper. The physics involved is too complex and even the evolution of the instability
itself is at present not known. The sole purpose of this section is to sketch the evolution
of the most important quantities of the process.

The instability proceeds in two steps: A phase characterized by an exponential growth, where
the linear approximation is valid, followed by a nonlinear phase. Contrary to the $r$-mode
instability, the ergoregion instability does not couple strongly to the fluid composing the
object. Therefore, the nonlinear phase is expected to be somewhat different from the
$r$-mode saturation phase \cite{Owen:1998xg,Arras:2002dw}.

As an illustration of waveform estimation, consider the instability triggered by a particle in
circular orbit around the compact object. Focusing on the $l=2$ mode, the metric perturbations
in the linear perturbation regime have the form
\beq
h_{+}&=& \frac{M}{D}h_0e^{t/\tau}\sin\left(\omega t-2\phi\right)\chi_+ \,,\\
h_{\times}&=& \frac{M}{D}h_0e^{t/\tau}\cos\left (\omega
t-2\phi\right )\chi_{\times}\,,
\label{waveform}
\eeq
where $D$ is the distance to the source, $h_0<<1$ and
\be
\chi_+\equiv \frac{\cos^2\theta+1}{2}\,,\quad
\chi_{\times}\equiv \cos\theta\,.\label{wave}
\ee
The above waveforms mimic the Newtonian waveform produced by a small mass in circular orbit
around the ultra-compact object. The linear perturbation regime corresponds to
$h_{+,\times}<1$. The exact functional form of the waveform (\ref{wave}) is required to
determine polarization and phase content of the waveform, but not the evolution of its
amplitude.

\begin{figure}[ht]
\begin{center}
\begin{tabular}{c}
\epsfig{file=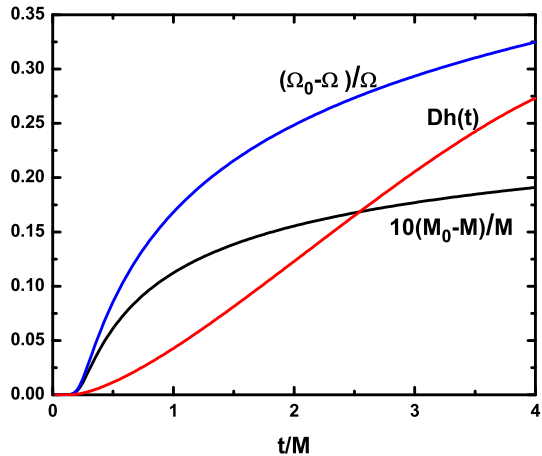,width=240pt,angle=0}
\end{tabular}
\end{center}
\caption{Evolution of mass, angular velocity and waveform of a thin-shell gravastar with
$h_0=10^{-4}$, $M\Omega_0=0.25$ and $\tau/M=0.041$ during the linear perturbation phase. The
mass difference is rescaled by a factor $10$ for better visualization.}
\label{fig:evolution}
\end{figure}

From the discussion in the previous sections, the frequency of the wave is $\omega\sim
\Omega(t)$. The timescale is $\tau\sim\tau_0(M\Omega)^{-6}$, where a dominant $w$-mode
instability is assumed \cite{Kokkotas:2002sf}. The mass can be determined as follows.  The
energy carried by the gravitational wave is \cite{teukolsky}
\be
\frac{d^2E}{dtd\Omega}=\lim_{D\rightarrow \infty}
\frac{D^2}{16\pi} \left (\dot{h}_+^2+\dot{h}_{\times}^2\right) \,.
\label{energywaves}
\ee
Assuming that all the energy carried by the gravitational wave is extracted from the star, it
follows $dE/dt=-dM/dt$. The angular momentum radiated in the azimuthal mode $m$ can be obtained
from $dE=\omega dJ/m$. If this angular momentum is also completely extracted from the star,
then $J=I\Omega$, where the moment of inertia $I=2\beta M(t)^3$ is approximately constant in
time. Setting $dJ/dt=m/\omega dE/dt=2/(\alpha\Omega)dM/dt$, the mass variation rate is
\be
\frac{\dot{M}}{M^3}=\alpha \beta \Omega\dot{\Omega}\,.
\label{Mdot}
\ee
Integration of Eq.~(\ref{Mdot}) yields
\be
\Omega^2=\Omega_0^2+\frac{M_0^{-2}-M^{-2}}{\alpha\beta}
\,.\label{solangularvel}
\ee
The mass and the angular velocity can be obtained by solving  the previous equations with the
condition $dE/dt=-dM/dt$. A typical solution is shown in Figure \ref{fig:evolution}.
%

\section{Discussion\label{sec:discussion}}
The above results show that ultra-compact objects with high redshift at their surface are
unstable when rapidly spinning. This strengthens the role of BHs as candidates for
astrophysical observations of rapidly spinning compact objects.

Boson stars and gravastars easily develop ergoregion instabilities. Analytical and numerical
results indicate that these objects are unstable against scalar field perturbations. Their
instability timescale is many orders of magnitude stronger than the instability timescale for
ordinary stars with uniform density \cite{cominsschutz}. In the large $l=m$ approximation,
suitable for a WKB treatment, gravitational and scalar perturbations have similar instability
timescales. In the low-$m$ regime gravitational perturbations are expected to have shorter
instability timescales than scalar perturbations.

For $J>0.4M^2$ instability timescales can be as low as a few tenths of a second for solar mass
objects and about a week for supermassive BHs, monotonically decreasing for larger rotations.
Therefore, high rotation is an indirect evidence for horizons. The spin of an astrophysical
compact object can be estimated by looking at the gas accreting near its surface
\cite{Narayan:2005ie, McClintock:2004, Fabian:2005hr}. A handful of fast-spinning BH-like
objects have been reported \cite{McClintock:2006xd, Wang:2006bz}. The results of this paper
suggest that these objects must indeed be BHs.

The ergoregion instability evolution is characterized by very distinct waveforms. If compact
astrophysical objects evolve through the ergoregion instability, gravitational-wave detectors
could easily identify them with a matched-filtering search.

It would be interesting to repeat the above analysis for other ultra-compact objects, such as
wormholes \cite{Morris:1988cz, visserbook, Lemos:2003jb,Damour:2007ap}, superspinars
\cite{Gimon:2007ur} and quark or fermion-boson stars \cite{fermionboson}. The general arguments
presented above suggest that these objects should be unstable. It would also be interesting to
perform a more detailed analysis of the gravitational sector of the ergoregion instability and
derive more refined templates for matched-filtering searches. A first step in this direction
would be to compute axial perturbations for any value of the azimuthal number $m$ along the
lines of previous works \cite{Kokkotas:2002sf}.
\section*{Acknowledgements}
The authors are indebted to Emanuele Berti for many useful
comments and suggestions, a critical reading of the manuscript and
for sharing some his numerical routines with us. The authors
warmly thank Burkhard Kleihaus, Jutta Kunz and Isabell Schaffer
for sharing and explaining their results concerning rotating boson
stars. They also thank \'Oscar Dias and Kostas Kokkotas for
interesting discussions and fruitful comments. This work was
partially funded by Funda\c c\~ao para a Ci\^encia e Tecnologia
(FCT) - Portugal through projects PTDC/FIS/64175/2006 and
POCI/FP/81915/2007. M.C. gratefully acknowledges the support of
the National Science Foundation through LIGO Research Support
Grant No. NSF PHY-0757937.

\end{document}